\documentclass[a4paper]{jpconf}
\usepackage{graphicx}
\usepackage{amsmath}
\usepackage{amsfonts}
\usepackage{amssymb}
\usepackage{xcolor}
\usepackage{hyperref}
\hypersetup{colorlinks=true, urlcolor=blue, linkcolor=black, citecolor=black, citebordercolor=green}
\DeclareMathOperator{\sech}{sech}
\usepackage{booktabs}
\usepackage{cite}
\usepackage{fancyhdr}
\begin{document}
\vspace*{-2.5cm}
\title{On varying coefficients of spatial inhomogeneous nonlinear Schr\"{o}dinger equation}
\author{N. Karjanto$^1$ and J. Tan$^{2,}$\footnote[3]{\!\!Present address: Frost \& Sullivan, 2, Jalan Kiara, Mont Kiara, Kuala Lumpur 50480, Malaysia.}}
\address{$^1$ Department of Mathematics, University College, Sungkyunkwan University, Natural Science Campus,
2066 Seobu-ro, Jangan-gu, Suwon 16419, Gyeonggi-do, Republic of Korea}
\address{$^2$ Department of Mechanical Engineering, Faculty of Engineering, The University of Nottingham Malaysia Campus, Semenyih 43500, Selangor, Malaysia}

\ead{natanael@skku.edu}

\thispagestyle{fancy}

\begin{abstract}
A nonlinear evolution equation for wave packet surface gravity waves with variation in topography is revisited in this article.
The equation is modeled by a spatial inhomogeneous nonlinear Schr\"{o}dinger (NLS) equation with varying coefficients,
derived by~Djordjevi\'{c} and Redekopp (1978) and the nonlinear coefficient is later corrected by Dingemans (1997).
We show analytically and qualitatively that the nonlinear coefficient and the corresponding averaging value, stated but not derived, by Benilov, Flanagan and Howlin (2005) and Benilov and Howlin (2006) are inaccurate. For a particular choice of topography and wave characteristics, the NLS equation alternates between focusing and defocusing case and hence, it does not admit the formation of a classical soliton, neither bright nor dark one.
\end{abstract}

\section{Introduction}
In this article, we revisit a nonlinear evolution equation for wave packet surface gravity waves with variation in topography.
The equation is described by the spatial nonlinear Schr\"{o}dinger (NLS) equation with varying coefficients.
It was derived in~\cite{djor78} and the correction to the nonlinear coefficient was given in~\cite{debn94,ding97,ding01}.
The equation has been studied in the context of smooth topography in~\cite{ben05} and strong smooth topography in~\cite{ben06}. 
However, it can be verified that the nonlinear coefficients used in these studies are inaccurate.
In this article, we compare analytically and qualitatively between the accurate and inaccurate nonlinear coefficients and a consequence of choosing 
particular wave characteristics and fluid depth variation to the model equation.

\section{On an evolution equation}
A nonlinear evolution for the packet of surface gravity waves propagating along uneven bottom is modeled by 
the spatial inhomogeneous nonlinear Schr\"{o}dinger (NLS) equation, also known as the NLS equation with varying, non-constant coefficients~\cite{djor78}.
We notice that the NLS equations with non-constant coefficients adopted in~\cite{ben05} and~\cite{ben06} presented in different expressions even though both model equations refer to~\cite{djor78}. Equation~(2.2) in~\cite{ben05} is written with negative signs in front of the dispersive and nonlinear coefficients, $\alpha$ and $\beta$, respectively. It reads
\begin{equation}
i \left(\partial_x A + c_g^{-1} \partial_t A + \mu A \right) - \alpha \partial_t^2 A - \beta A |A|^2 = 0.
\end{equation}
Meanwhile, equation~(6) in~\cite{ben06} is written with positive signs in front of the dispersive and nonlinear coefficients, $\alpha(x)$ and $\beta(x)$, respectively. It reads
\begin{equation}
i \left(\partial_x A + c_g(x)^{-1} \partial_t A + \mu(x) A \right) + \alpha(x) \partial_t^2 A + \beta(x) A |A|^2 = 0.
\end{equation}
For both NLS equations, $A(x,t)$ refers to the complex-valued amplitude of the wave packet for surface gravity waves propagating along topography variation,
$c_g$ is the group velocity and $\mu$ is called a dissipative coefficient; also known as an amplification/absorption coefficient in nonlinear optics; a chemical potential in Bose-Einstein condensates. It is obvious that both dispersive and nonlinear coefficients $\alpha(x)$ and $\beta(x)$ have the opposite sign from one paper to another.
However, these coefficients are expressed identically in both papers, with the exception of writing $h$ and $H$ for the fluid depth and 
the term $k^2 c_g \sech^2 kH$ expressed in an equivalent form $-c_g(\omega^4 - k^2)$ using a hyperbolic functions relationship $\sech^2 kH = 1 - \tanh^2(kH)$. The dispersive and nonlinear coefficients are presented as follows, taken from~\cite{ben06}, where later in our notation we 
display them using the normalized quantities $\tilde{k}$ and $\tilde{\omega}$:
\vspace*{-0.1cm}
\begin{align}
\alpha(x) &= \frac{1}{2\omega c_g} \left(1 - \frac{H}{c_g^2} + \frac{2\omega H \tanh kH}{c_g} \right) \label{alphax} \\
\beta(x)  &= \frac{1}{2 \omega^3 c_g} \left(3k^4 + 2 \omega^4 k^2 - \omega^8 - \frac{(2k \omega + k^2 c_g \sech^2 kH)^2}{H - c_g^2} \right). \label{betaben}
\end{align}
The dispersive coefficient $\alpha(x)$ should be the negative of expression~\eqref{alphax}, hence, the expression in~\cite{ben05} is correct, up to a scaling factor of the constant gravitational acceleration $g$. The nonlinear coefficient $\beta(x)$~\eqref{betaben} is found to be inaccurate and the accurate one is given explicitly in Section~\ref{disnoncoeff} and shown the accuracy qualitatively.
\vspace*{-0.5cm}
\begin{figure}[h]
\begin{center}
\includegraphics[width = 0.35\textwidth]{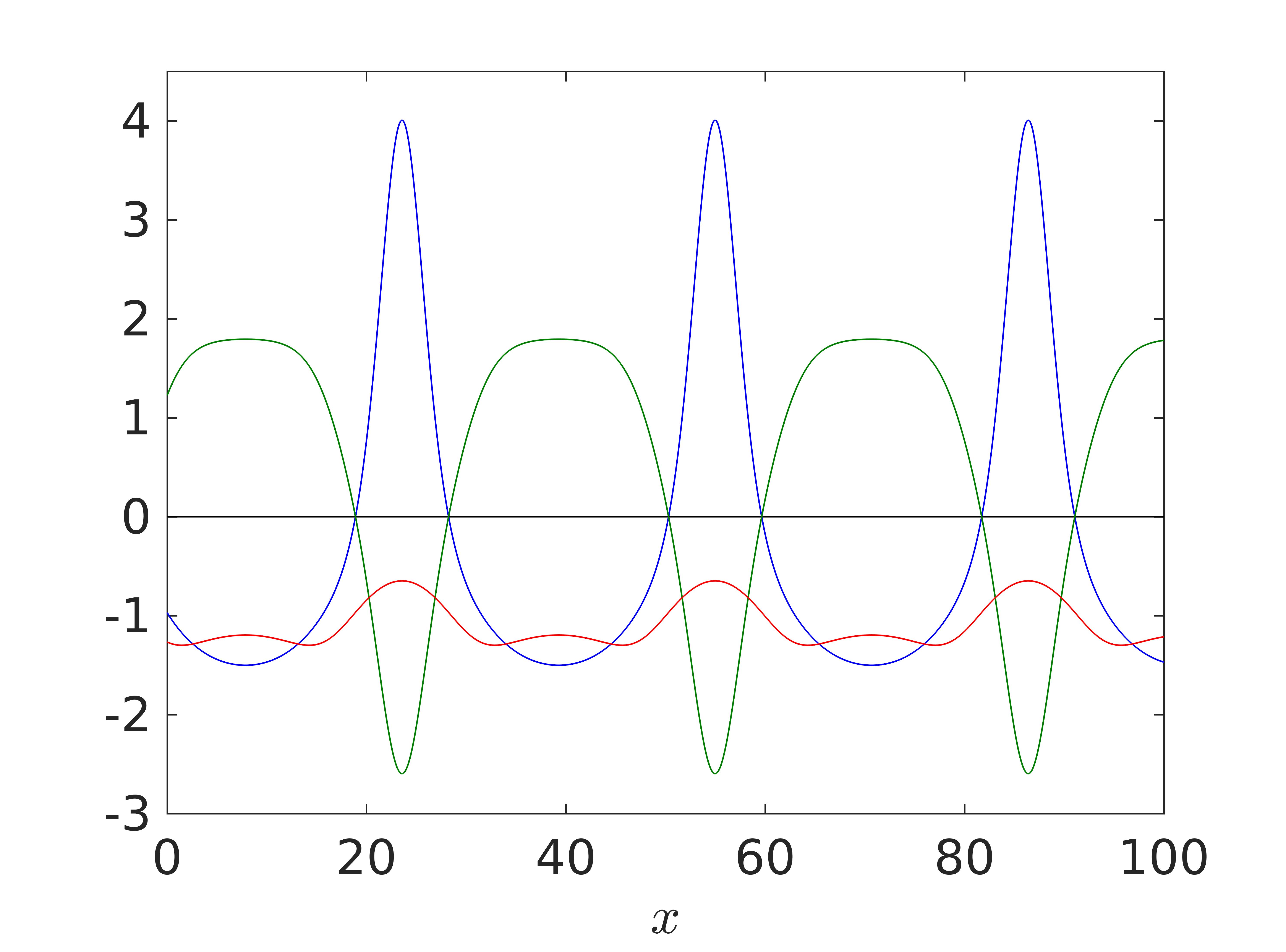} \hspace{2cm}
\includegraphics[width = 0.35\textwidth]{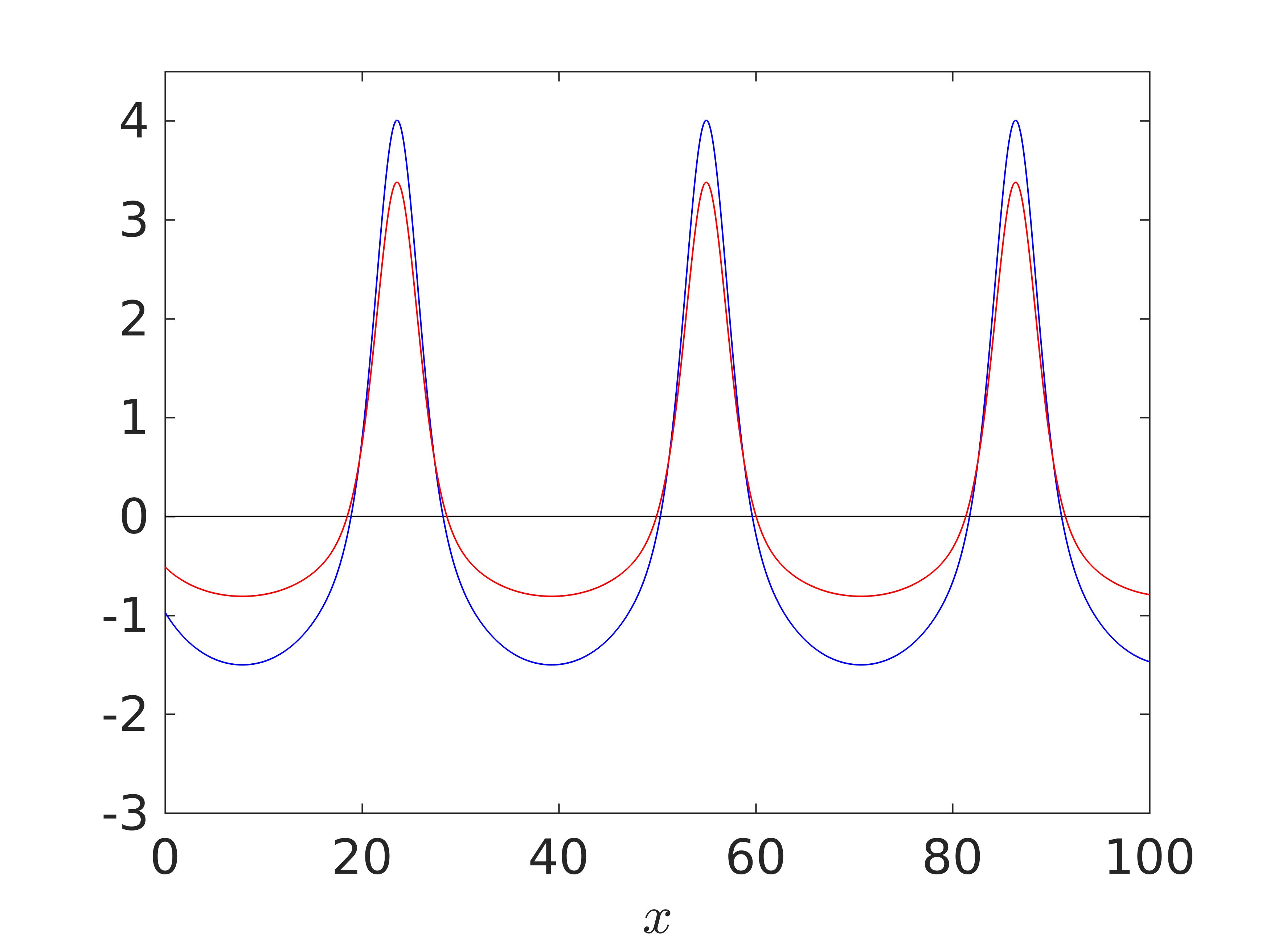}
\end{center}
\vspace*{-0.5cm}
\caption{\small (Left panel) Plots of the dispersive coefficient $\alpha(x)$ (red), the (scaled) nonlinear coefficient $\beta(x)$ (blue) and
the product of the dispersive and the nonlinear coefficient $\alpha(x) \beta(x)$ (green). We observe that the NLS equation alternates between focusing and defocusing case along the spatial evolution parameter $x$.
(Right panel) Comparison plots of the (scaled) nonlinear coefficients $\beta(x)$ between an accurate one derived in~\cite{djor78} and corrected in~\cite{ding97, ding01} (blue curve) and an inaccurate one (red curve), stated in~\cite{ben05} and~\cite{ben06} by referring to~\cite{djor78}, but not derived.
All cases correspond to the water depth $H(x) = 1 + 0.6 \sin (x/5)$.} \label{alphabetaH412}
\end{figure}

Applying a multiple scale method in a co-moving reference by replacing time $t$ with ${\displaystyle \tau = t - \int c_g(x)^{-1} dx}$ 
and writing the coefficient $\mu$ as ${\displaystyle c_g(x)^{-1} c_g'(x)}$, the NLS equation can be written as follows, where we adopt the positive sign of the coefficients in the notation
\vspace*{-0.1cm}
\begin{equation}
i \partial_x \left(\sqrt{c_g} A \right) + \alpha(x) \partial_\tau^2 \left(\sqrt{c_g} A \right) + c_g(x)^{-1}\beta(x) \left|\sqrt{c_g} A \right|^2 \left(\sqrt{c_g} A \right) = 0. \label{iNLSpos}
\end{equation}
There exists a relationship between the NLS equation in this form with the one expressed with negative signs in front of the dispersive and nonlinear coefficients.
Employing the transformation $A = - \tilde{A}^\ast$, where the $\ast$ denotes the complex conjugate, we obtain a similar NLS equation for $\tilde{A}$, with negative signs in front of the dispersive and nonlinear coefficients, and yet these varying coefficients are identical with ones in~\eqref{iNLSpos}, given as follows:
\vspace*{-0.1cm}
\begin{equation}
i \partial_x \left(\sqrt{c_g} \tilde{A} \right) - \alpha(x) \partial_\tau^2 \left(\sqrt{c_g} \tilde{A} \right) - c_g(x)^{-1}\beta(x) \left|\sqrt{c_g} \tilde{A} \right|^2 \left(\sqrt{c_g} \tilde{A} \right) = 0. \label{iNLSneg}
\end{equation}
The difference in the solutions for the NLS equation~\eqref{iNLSpos} and~\eqref{iNLSneg} is only in the phase shift. Indeed, if
${\displaystyle A(x,\tau) = a(x,\tau) e^{i\phi(x,\tau)}}$ is a solution of the NLS equation~\eqref{iNLSpos} in its phase-amplitude form, then
${\displaystyle \tilde{A}(x,\tau) = a(x,\tau) e^{i\left[\pi - \phi(x,\tau) \right]}}$
is a solution for the NLS equation~\eqref{iNLSneg}.
 
\section{On dispersive and nonlinear coefficients} \label{disnoncoeff}
In the following, we adopt the spatial inhomogeneous NLS equation with positive signs of dispersive and nonlinear coefficients
and hence both coefficients carry the negative sign in front of them.
The original dispersive coefficient derived in~\cite{djor78} reads
\vspace*{-0.2cm}
\begin{equation}
\alpha(x) = \frac{-1}{2\omega c_g} \left(1 - \frac{gh}{c_g^2} (1 - kh \sigma)(1 - \sigma^2) \right), \qquad \text{where} \qquad \sigma = \tanh kh.
\vspace*{-0.2cm}
\end{equation}
It can be shown that the normalized dispersive coefficient $\tilde{\alpha}(x)$ can be written as follows, which is in agreement with the expression~\eqref{alphax}, presented in~\cite{ben05} and~\cite{ben06}, where the latter one should include the negative sign:
\vspace*{-0.4cm}
\begin{equation}
\tilde{\alpha}(x) = \frac{\alpha(x)}{g} = \frac{-1}{2 \tilde{\omega} c_g} \left(1 - \frac{H}{c_g^2} + \frac{2 \tilde{\omega} H \tanh \tilde{k}H}{c_g} \right).
\vspace*{-0.2cm}
\end{equation}
Here, the physical water depth is scaled to normalized depth $H(x)$ using the relation $gh(x) = H(x)$.
It follows that the normalized wave frequency $\tilde{\omega}$ and the normalized wavenumber $\tilde{k}$ 
are given by the following relations: $\omega^2 = g^2 \tilde{\omega}^2$ and $k = g\tilde{k}$, respectively. 
The plot of the dispersive coefficient $\alpha(x)$ is displayed as the red curve on the left panel of Figure~\ref{alphabetaH412}.
\vspace*{-0.3cm}
\begin{figure}[h]
\begin{minipage}{20pc}
\includegraphics[width=20pc]{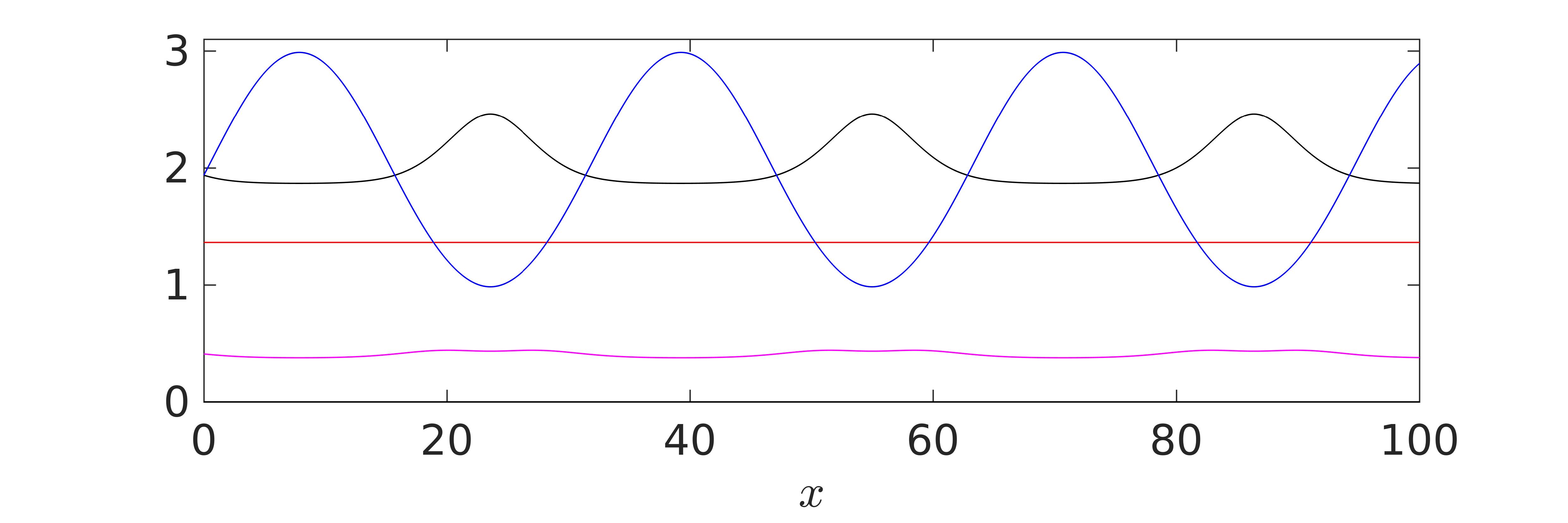}
\vspace*{-0.5cm}
\caption{\small \label{kcgomega412a} Plots of wave frequency $\tilde{\omega} = 1.363 = kh$, depicted as a red constant line, wavenumber $k(x)$ (black), 
the product of wavenumber and the water depth $kh$ (blue) and the group velocity $c_g$ (magenta).}
\end{minipage}\hspace{1pc}%
\begin{minipage}{17pc}
\includegraphics[width=17pc]{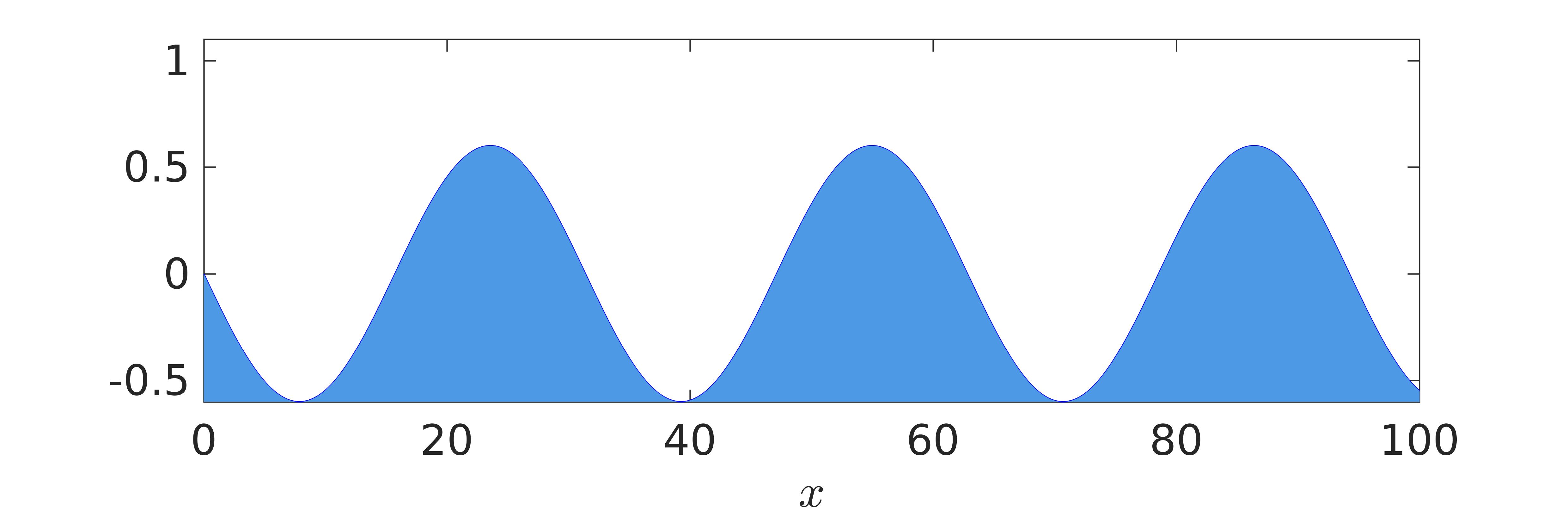}
\vspace*{-0.5cm}
\caption{\small \label{kcgomega412b} Sketch of the bottom topography corresponds to the water depth $H(x) = 1 + 0.6 \sin (x/5)$.}
\end{minipage} 
\end{figure}

It is noted in~\cite{ding97, ding01} that there exists a tiny typographical error in the nonlinear coefficient given in~\cite{djor78},
the expression of the final term should be $(1 - \sigma^2)^2$ instead of $(1 - \sigma)^2$.
The original expression of the accurate nonlinear coefficient reads, where the phase velocity $c_p = \omega/k$:
\vspace*{-0.2cm}
\begin{equation}
\beta(x) = \frac{-k^4}{4 \omega \sigma^2 c_g} \left[ 9 - 10 \sigma^2 + 9 \sigma^4  
- \frac{2\sigma^2 c_g^2}{gh - c_g^2} \left\{4 \left(\frac{c_p}{c_g} \right)^2 + 4 \frac{c_p}{c_g} (1 - \sigma^2) + \frac{gh}{c_g^2} (1 - \sigma^2)^2  \right\} \right].
\vspace*{-0.2cm}
\end{equation}
After some manipulation, the normalized nonlinear coefficient $\tilde{\beta}$ can be expressed as follows:
\vspace*{-0.2cm}
\begin{equation}
\tilde{\beta}(x) = \frac{\beta(x)}{g^3} = \frac{-1}{4 \tilde{\omega}^5 c_g} \left[ 9 \tilde{k}^6 - 12 \tilde{k}^4 \tilde{\omega}^4  + 13 \tilde{k}^2 \tilde{\omega}^8 - 2 \tilde{\omega}^{12}\right] 
 + \frac{1}{2 \tilde{\omega} c_g} \frac{\left[2 \tilde{k} \tilde{\omega} + \tilde{k}^2 c_g \sech^2 \tilde{k}H \right]^2}{H - c_g^2}. \label{beta78} 
\end{equation}
We observe that this nonlinear coefficient exhibits a different expression with the one presented and used in~\cite{ben05} and~\cite{ben06}. Compare with~\eqref{betaben}. The plot of accurate nonlinear coefficient is displayed as a blue curve on the left panel of Figure~\ref{alphabetaH412}.
The comparison between an accurate $\beta(x)$ and an inaccurate one for a particular water depth is displayed on the right panel of Figure~\ref{alphabetaH412}. 
The left panel of Figure~\ref{alphabetaH412} also shows the plot of the product of the dispersive and the nonlinear coefficients
$\alpha(x) \beta(x)$, indicating that the NLS equation alternates between focusing and defocusing case along the spatial evolution parameter $x$.
In the following, we show qualitatively that the nonlinear coefficient presented in~\cite{ben05} and~\cite{ben06} is indeed inaccurate.

Note that the nonlinear coefficient $\beta$ changes sign when $kh \approx 1.363$~\cite{ben67}.
We observe in Figure~\ref{kcgomega412a} that for $kh > 1.363$, the carrier wave is unstable with respect to modulation and hence bright, single-soliton solution is formed. On the other hand, for $kh < 1.363$, the NLS equation does not possess bright soliton solution, but dark soliton instead since the carrier wave is modulationally stable. For this example, the scaled water depth is given by $H(x) = 1 + 0.6 \sin(x/5)$ and the bottom topography is depicted in Figure~\ref{kcgomega412b}.

\begin{figure}[h]
\begin{center}
\includegraphics[width = 0.35\textwidth]{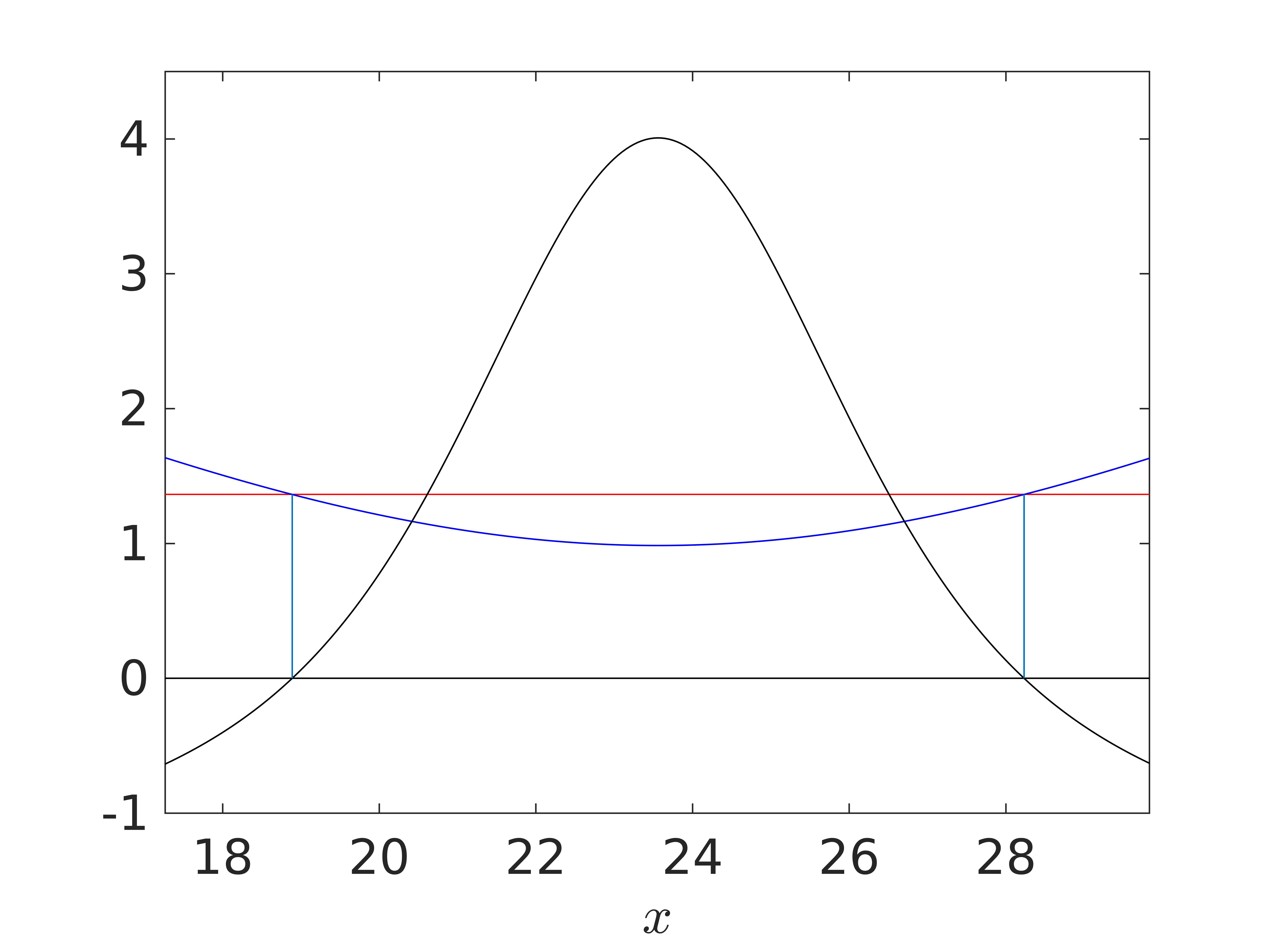} \hspace{2cm}
\includegraphics[width = 0.35\textwidth]{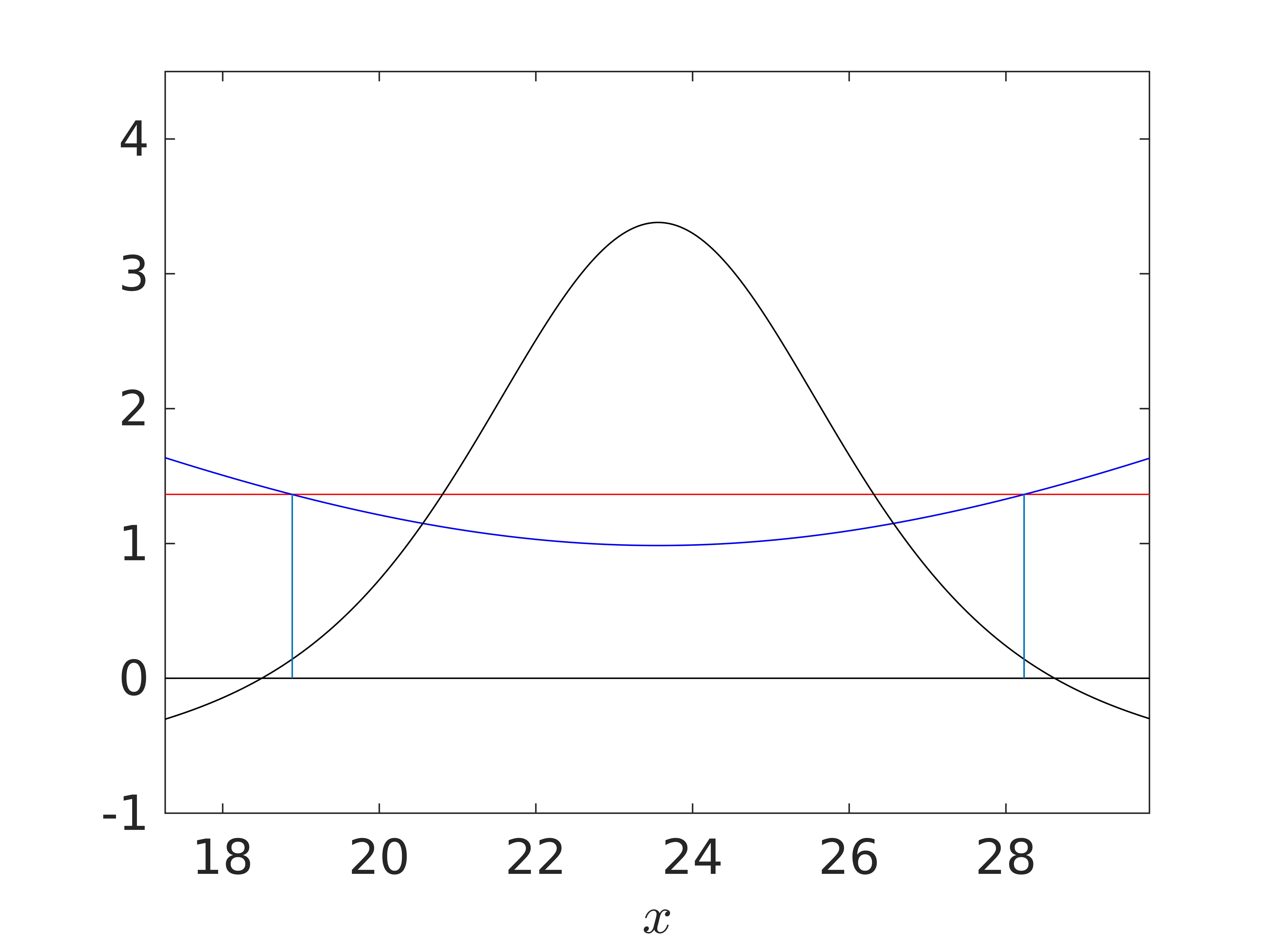}
\end{center}
\vspace*{-0.5cm}
\caption{\small (Left panel) Plots of wave frequency $\tilde{\omega} = 1.363 = kh$, depicted as a red constant line, 
the product of wavenumber and the water depth $kh$ (blue) and the nonlinear coefficient $\beta$ (black), after some scaling.
(Right panel) Similar to the left panel except for the inaccurate nonlinear coefficient $\beta$ stated in~\cite{ben05}.
We observed that when $kh \approx 1.363$, it does not correspond to the change of sign in the nonlinear coefficient.
All cases correspond to the water depth $H(x) = 1 + 0.6 \sin (x/5)$.} \label{kH412beta}
\end{figure}

The change of signs for the nonlinear coefficient corresponds to the change of the value in $kh$, from $kh > 1.363$ to $kh < 1.363$, or vice versa.
Figure~\ref{kH412beta} shows zoom-in plots of $kh$ and the scaled nonlinear coefficient $\beta$ for the wave frequency $\tilde{\omega} = 1.363$.
The left panel displays the accurate nonlinear coefficient derived in~\cite{djor78} and corrected in~\cite{ding97, ding01} while 
the right panel displays inaccurate nonlinear coefficient given in~\cite{ben05} and~\cite{ben06} by referring to~\cite{djor78}, with no given derivation.
For both cases, the corresponding water depth is $H(x) = 1 + 0.6 \sin(x/5)$.
In the left panel, we observe that when $kh > 1.363$, the nonlinear coefficient $\beta < 0$, and since $\alpha < 0$, this gives a focusing case for the NLS equation
and the wave train is modulationally unstable. 
When $kh < 1.363$, the nonlinear coefficient $\beta > 0$ and since $\alpha > 0$, the NLS equation is defocusing type and the wave train is modulationally stable.
The situation does not occur for an inaccurate nonlinear coefficient, as can be observed in the right panel of Figure~\ref{kH412beta}. 
\vspace*{-0.4cm}
\vspace*{-0.5cm}

\section{On averaging coefficients}
For a particular choice of initial wavenumber $k(x = 0) = 2$, which corresponds to the wave frequency $\omega = 1.3885$,
and the the water depth $H(x) = 1 + 0.6 \sin (x/5)$, the NLS equation alternates between focusing and defocusing case along the spatial evolution parameter $x$.
This is due to the fact that the product of the dispersive and the nonlinear coefficients $\alpha(x) \beta(x)$ is alternating between positive and negative along the horizontal position $x$.
Hence, the formation of only bright, single-soliton solution is not admitted.
Nonetheless, both asymptotic and numerical solutions can be found and a good comparison of the relative amplitude of the wave packet between the two results is presented. See Figure~2(a) in~\cite{ben06}.
The asymptotic solution is presented in term of averaging value of the dispersive and nonlinear coefficients as well as the initial condition for the numerical computation. The averaging of a function $f(x)$ with respect to $x$ over one period of topography $T$ is defined as follows:
\begin{equation}
\langle f \rangle = \frac{1}{T} \int_{0}^{T} f(x) \, dx.
\end{equation}

Using this definition, the averaging value of the dispersion and nonlinear coefficients are readily calculated.
For the choice of wave frequency, initial wavenumber and the water depth mentioned above, the period of topography $T = 10\pi$ and we obtain $\langle \alpha \rangle = 1.0868$ and for the wave amplitude $\eta = 0.1$, the scale of dispersion $L_d = 92.0158$. However, a disagreement is found for the scale of nonlinearity $L_n$.
Our calculation shows $L_n = 276.026$ while in~\cite{ben06}, $L_n = 114.1$.
This is due to the difference of the averaging value for the nonlinear coefficient $c_g^{-1} \beta$.
Our calculation finds $\langle c_g^{-1} \beta \rangle = 2.2164$ using the inaccurate $\beta$ while in~\cite{ben06}, $\langle c_g^{-1} \beta \rangle = 11.627$.
Using the accurate value of $\beta$, the averaging value $\langle c_g^{-1} \beta \rangle = -22.4010$.
A comparison of the averaging values for several wave frequencies and initial wavenumbers are displayed in Table~\ref{compare}.
\vspace*{-0.5cm}
\begin{table}[h]
\centering
\caption{\label{compare} A comparison of the averaging values of the nonlinear coefficient $\beta/c_g$.
The signs in $\langle c_g^{-1} \beta \rangle$ from~\cite{ben06} and the calculation using inaccurate $\beta$ from~\cite{ben06} are not adjusted 
(the third and the fourth columns). The sign in $\langle c_g^{-1} \beta \rangle$ has been corrected (the fifth column).} 
\begin{tabular}{@{}ccrrr@{}}	
\br 
Initial       & Wave               & $\langle c_g^{-1} \beta \rangle$  & $\langle c_g^{-1} \beta \rangle$  & $\langle c_g^{-1} \beta \rangle$ \\
wavenumber    & frequency          & from~\cite{ben06}                 & (inaccurate $\beta$)              & (accurate $\beta$) \\ 
\mr 
$k(0) = 2.0$  & $\omega = 1.3885$  & 11.627   & $ 2.2164$  & $-22.4010$  \\ 
$k(0) = 1.9$  & $\omega = 1.3479$  &  3.720   & $-6.2451$  & $-6.7465$   \\ 
$k(0) = 1.8$  & $\omega = 1.3055$  & $-2.906$ & $-13.4684$ & $4.9603$   \\ 
\br 
\end{tabular}
\end{table} 

\vspace*{-0.5cm}
\section{Conclusion}
We revisited the spatial nonlinear Schr\"{o}dinger equation with non-constant coefficients as a nonlinear evolution equation for wave packet of surface gravity waves when it propagates over an uneven bottom, i.e. a topography with some variation. We have rewritten the dispersive and nonlinear coefficients in the form that easier to compare with the ones found in the literature. We have shown qualitatively using the change of values in $kh$ and the change in its sign that the nonlinear coefficient used in~\cite{ben05} and~\cite{ben06} are inaccurate. A similar pattern of this inaccuracy is also carried along in the averaging value computation of the nonlinear coefficient used for an asymptotic solution of the transformed NLS equation. For a particular choice of wave characteristics and water depth, the NLS equation alternates between focusing and defocusing case along the spatial evolution parameter. Hence, the formation of the classical soliton, either bright or dark, is not admitted for this case.

\ack
{\small The authors would like to thank Andy Chan, Siew Ling Ng and Margaret Rozario (The University of Nottingham Malaysia Campus), 
Teo Lee Peng (Xiamen University Malaysia Campus), Juan-Ming Yuan (Providence University, Taiwan), Mason Porter (UCLA) and
Brenny van Groesen (Universiteit Twente, The Netherlands and LabMath Indonesia) for assistance, fruitful discussions and suggestions.
The New Researcher Fund from The University of Nottingham University Park UK Campus and Malaysia Campus NRF 5035-A2RL20,
Samsung Intramural Research Fund Project 2016-1299-000 and National Research Foundation of Korea Fund NRF-2017017743 are gratefully acknowledged. \par}

\newpage
\section*{Appendix} 
This appendix provides a technical detail on derivations of the varying coefficients for the NLS equation with non-constant coefficients,
for which different source of literature expresses them differently. 
The purpose is bridging the gap between~\cite{djor78,ding01} and~\cite{ben05,ben06}.
We will consider the dissipative, dispersive and nonlinear coefficients, $\mu(x)$, $\alpha(x)$ and $\beta(x)$, respectively.

To begin with, we have a relationship between the wave frequency $\omega$ and the wavenumber $k$ by the linear dispersion relation.
Using the transformation $gh = H$, we can normalized both the wavenumber and wave frequency as $(\tilde{k}, \tilde{\omega})$
using the relationship $k = g \tilde{k}$ and $\omega^2 = g^2 \tilde{\omega}$.
The original dispersion relation and its scaled version are given as follows (note that $kh = \tilde{k} H$):
\begin{equation}
\omega^2 = gk \sigma = gk \tanh kh \qquad \qquad \qquad \qquad \tilde{\omega}^2 = \frac{\omega^2}{g^2} = \tilde{k} \sigma = \tilde{k} \tanh \tilde{k} H.
\label{ldrkua}
\end{equation}
The group velocity $c_g$ is defined as follows and it can be written in normalized quantities:
\begin{equation}
c_g \equiv \frac{\partial \omega}{\partial k} = \frac{g}{2 \omega} \left[\sigma + kh (1 - \sigma^2) \right] \qquad \qquad
c_g(x) = \frac{1}{2 \tilde{\omega}} \left[\sigma + \tilde{k}H (1 - \sigma^2) \right]. \label{grovel}
\end{equation}
Note that the expression for $c_g$, between expressions (2.8) and (2.9) in~\cite{djor78}, is not $\Omega$, but $\omega$ instead.
A similar typographical error is also found in expression (2.16) of~\cite{djor78}. See expression~\eqref{disco}. 

The dissipative coefficient $\mu$ in the field of nonlinear optics it is also known as an amplification or absorption coefficient,
while in Bose-Einstein condensation, it is often called as the chemical potential. An explicit expression is given as follows and it can also be expressed in a simpler form in term of the group velocity $c_g$:
\begin{equation}
\mu(x) = -\frac{k'(x)}{k(x)} \frac{\sigma(1 - kh \sigma)}{\sigma + kh (1 - \sigma^2)} 
= \frac{(1 - \sigma^2)(1 - kh \sigma)}{\sigma + kh (1 - \sigma^2)} \frac{d(kh)}{dx} = \frac{1}{c_g} \frac{dc_g}{dx}. \label{mu}
\end{equation}
To show the second equality in~\eqref{mu}, we need simply need to verify that 
\begin{equation}
-\frac{k'(x)}{k(x)} \sigma = (1 - \sigma^2) \frac{d(kh)}{dx}. \label{mu0}
\end{equation}
This can be shown using the fact that the wave frequency $\omega$ is constant along the spatial wave propagation $x$.
Hence,
\begin{equation}
\frac{d}{dx} \left(\frac{\omega^2}{g} \right) = \frac{d}{dx} (k\sigma) = k'(x) \sigma + k \sigma'(x) = 0
\end{equation}
which implies that~\eqref{mu0} is easily obtained
\begin{equation}
-\frac{k'(x)}{k(x)} \sigma = \sigma'(x) = \frac{d}{dx} \tanh kh = \sech^2 kh \cdot \frac{d}{dx}(kh) = (1 - \sigma^2) \frac{d (kh)}{dx}.
\end{equation}
The proof of the third equality in~\eqref{mu} is given as follows:
\begin{align}
\mu(x) &= \frac{(1 - \sigma^2)(1 - kh \sigma)}{\sigma + kh (1 - \sigma^2)} \frac{d(kh)}{dx} 
= \frac{2g(1 - \sigma^2)(1 - kh \sigma)}{2g \left[\sigma + kh (1 - \sigma^2) \right]} \frac{d(kh)}{dx} \nonumber \\
&= \frac{g}{4 \omega c_g}  \left[ 2 (1 - \sigma^2) - 2(1 - \sigma^2) kh \sigma \right] \frac{d(kh)}{dx} 
= \frac{g}{4 \omega c_g} \left[ \sigma'(x) + (1 - \sigma^2) + kh (- 2 \sigma \sigma') \right] \frac{d(kh)}{dx} \nonumber \\
\mu(x) &= \frac{1}{4 \omega c_g} \frac{d}{dx} (2 \omega c_g) = \frac{1}{2 c_g} \frac{d c_g}{dx}, \qquad \qquad \text{since} \quad \omega = \text{constant}.
\end{align}

The dispersive coefficient $\alpha$ is given as follows
\begin{equation}
\alpha(x) \equiv \frac{1}{2 c_g^3} \frac{\partial^2 \omega}{\partial k^2} = \frac{-1}{2\omega c_g} \left(1 - \frac{gh}{c_g^2} (1 - kh \sigma)(1 - \sigma^2) \right).
\label{disco}
\end{equation}
To show this, we start by taking the partial derivative of $\omega^2$~\eqref{ldrkua} with respect to $k$
\begin{equation}
\frac{\partial \omega^2}{\partial k} = 2 \omega \frac{\partial \omega}{\partial k} = g \sigma + g k \frac{\partial \sigma}{\partial k}
= g \sigma + gkh (1 - \sigma^2). \label{tuomkua}
\end{equation}
We observe that from~\eqref{tuomkua}, an expression for the group velocity $c_g$~\eqref{grovel} is obtained.
Taking the partial derivative again of $\omega^2$~\eqref{ldrkua} with respect to $k$ yields
\begin{align}
\frac{\partial^2 \omega^2}{\partial k^2} &= 2 \left(\frac{\partial \omega}{\partial k}\right)^2 + 2 \omega \frac{\partial^2 \omega}{\partial k^2}
= 2 c_g^2 + 2 \omega \frac{\partial^2 \omega}{\partial k^2} \label{tuduomkua1} \\
&= 2 g \frac{\partial \sigma}{\partial k} + g k \frac{\partial^2 \sigma}{\partial k^2} = 2 gh (1 - \sigma^2) - 2 gh (1 - \sigma^2) kh \sigma
= 2 gh (1 - kh \sigma) (1 - \sigma^2). \label{tuduomkua2}
\end{align}
Combining the final expressions in~\eqref{tuduomkua1} and~\eqref{tuduomkua2}, we obtain the dispersive coefficient $\alpha(x)$~\eqref{disco}.
Furthermore, the dispersive coefficient $\alpha$ can also be re-written in the following form, and $\tilde{\alpha}$ is the corresponding normalized quantity.
\begin{align}
\alpha(x) &= \frac{-1}{2\omega c_g} \left(1 - \frac{gh}{c_g^2} (1 - kh \sigma)(1 - \sigma^2) \right) 
= \frac{-1}{2\omega c_g} \left(1 - \frac{gh}{c_g^2} (1 - \sigma^2 - kh \sigma + kh \sigma^3) \right) \nonumber \\
&= \frac{-1}{2\omega c_g} \left(1 - \frac{gh}{c_g^2} + \sigma \frac{gh}{c_g^2} \left[ \sigma + kh (1 - \sigma^2) \right] \right) 
= \frac{-1}{2\omega c_g} \left(1 - \frac{gh}{c_g^2} + \sigma \frac{gh}{c_g^2} \cdot \frac{2 \omega c_g}{g} \right) \nonumber \\
&= \frac{-1}{2\omega c_g} \left(1 - \frac{gh}{c_g^2} + \frac{2 \omega h \sigma}{c_g} \right) 
= \frac{-1}{2\omega c_g} \left(1 - \frac{gh}{c_g^2} + \frac{2\omega h \tanh kh}{c_g} \right) \nonumber \\
\tilde{\alpha}(x) = \frac{\alpha(x)}{g} &= \frac{-1}{2 \tilde{\omega} c_g} \left(1 - \frac{H}{c_g^2} + \frac{2 \tilde{\omega} H \tanh \tilde{k}H}{c_g} \right).
\end{align}

Finally, we show the derivation of the nonlinear coefficient $\beta$. Here, $c_p = \omega/k$ is the phase velocity.
\begin{align}
\beta(x) &= \frac{-k^4}{4 \omega \sigma^2 c_g} \left[ 9 - 10 \sigma^2 + 9 \sigma^4  
- \frac{2\sigma^2 c_g^2}{gh - c_g^2} \left\{4 \left(\frac{c_p}{c_g} \right)^2 + 4 \frac{c_p}{c_g} (1 - \sigma^2) + \frac{gh}{c_g^2} (1 - \sigma^2)^2  \right\} \right] \nonumber \\
&= \frac{-k^4}{4 \omega \sigma^2 c_g} \left( 9 - 10 \sigma^2 + 9 \sigma^4 \right) +  
\frac{k^4 c_g^2}{2 \omega c_g (gh - c_g^2)} \left[ \left(2 \frac{c_p}{c_g} + (1 - \sigma^2) \right)^2 + \left(\frac{gh}{c_g^2} - 1 \right) (1 - \sigma^2)^2 \right] 
\nonumber \\
&=\text{\small $\frac{-k^4}{4 \omega \sigma^2 c_g} \left[ 9\left( 1 - 2 \sigma^2 + \sigma^4 \right) + 8 \sigma^2 \right] +  
\frac{k^4 c_g^2}{2 \omega c_g (gh - c_g^2)} \left[ \frac{\left[2 c_p + c_g (1 - \sigma^2) \right]^2}{c_g^2} + \left(\frac{gh - c_g^2}{c_g^2} \right) (1 - \sigma^2)^2 \right]$} \nonumber \\
&= \frac{-k^4}{4 \omega \sigma^2 c_g} \left[ 9\left( 1 - \sigma^2 \right)^2 + 8 \sigma^2 - 2\sigma^2 (1 - \sigma^2)^2 \right] +  
\frac{1}{2 \omega c_g} \frac{\left[2 c_p k^2 + k^2 c_g (1 - \sigma^2)\right]^2}{gh - c_g^2} \nonumber \\
&= \frac{-k^4}{4 \omega \sigma^2 c_g} \left\{ 9\left( 1 - \sigma^2 \right)^2 + 2 \sigma^2 \left[ 4 - (1 - \sigma^2)^2 \right]  \right\} +  
\frac{1}{2 \omega c_g} \frac{\left[2 k \omega + k^2 c_g (1 - \tanh^2 kh)\right]^2}{gh - c_g^2} \nonumber \\
&= \frac{-k^4}{4 \omega \sigma^2 c_g} \left\{ 9\left( 1 - \sigma^2 \right)^2 + 2 \sigma^2 \left[ 2 - (1 - \sigma^2) \right] \left[ 2 + (1 - \sigma^2) \right] \right\} +  \frac{1}{2 \omega c_g} \frac{\left[2 k \omega + k^2 c_g \sech^2 kh\right]^2}{gh - c_g^2} \nonumber \\
\beta(x) &= \frac{-k^4}{4 \omega \sigma^2 c_g} \left[ 9\left( 1 - \sigma^2 \right)^2 + 2 \sigma^2 \left(1 + \sigma^2 \right) ( 3 - \sigma^2) \right] + \frac{1}{2 \omega c_g} \frac{\left[2 k \omega + k^2 c_g \sech^2 kh\right]^2}{gh - c_g^2}
\end{align}
Since $\tilde{\omega}^2 = \tilde{k} \sigma$, then we can write an expression for the normalized nonlinear coefficient $\tilde{\beta}$:
{\small 
\begin{align}
\tilde{\beta}(x) &= \frac{\beta(x)}{g^3} = \frac{-\tilde{k}^6}{4 \tilde{\omega}^5 c_g} \left[ 9\left( 1 - \sigma^2 \right)^2 + 2 \sigma^2 \left(1 + \sigma^2 \right) ( 3 - \sigma^2) \right] + \frac{1}{2 \tilde{\omega} c_g} \frac{\left[2 \tilde{k} \tilde{\omega} + \tilde{k}^2 c_g \sech^2 \tilde{k}H \right]^2}{H - c_g^2} \nonumber \\
&= \frac{-1}{4 \tilde{\omega}^5 c_g} \left[ 9 \tilde{k}^2 \left( \tilde{k}^4  - 2 \tilde{k}^2 (\tilde{k}^2 \sigma^2) + \tilde{k}^4 \sigma^4 \right) 
+ 2 \tilde{k}^2 \sigma^2 \left(3 \tilde{k}^4 + 2 \tilde{k}^2 (\tilde{k}^2 \sigma^2) - \tilde{k}^4 \sigma^4 \right) \right] \nonumber \\
& \; + \frac{1}{2 \tilde{\omega} c_g} \frac{\left[2 \tilde{k} \tilde{\omega} + \tilde{k}^2 c_g \sech^2 \tilde{k}H \right]^2}{H - c_g^2} \nonumber \\
&= \frac{-1}{4 \tilde{\omega}^5 c_g} \left[ 9 \tilde{k}^2 \left( \tilde{k}^4  - 2 \tilde{k}^2 \tilde{\omega}^4  + \tilde{\omega}^8 \right) 
 + 2 \tilde{\omega}^4 \left(3 \tilde{k}^4 + 2 \tilde{k}^2 \tilde{\omega}^4 - \tilde{\omega}^8 \right) \right] 
 + \frac{1}{2 \tilde{\omega} c_g} \frac{\left[2 \tilde{k} \tilde{\omega} + \tilde{k}^2 c_g \sech^2 \tilde{k}H \right]^2}{H - c_g^2} \nonumber \\
&= \frac{-1}{4 \tilde{\omega}^5 c_g} \left[ 9 \tilde{k}^6 - 12 \tilde{k}^4 \tilde{\omega}^4  + 13 \tilde{k}^2 \tilde{\omega}^8 - 2 \tilde{\omega}^{12}\right] 
 + \frac{1}{2 \tilde{\omega} c_g} \frac{\left[2 \tilde{k} \tilde{\omega} + \tilde{k}^2 c_g \sech^2 \tilde{k}H \right]^2}{H - c_g^2} \\
\tilde{\beta}(x) &= \frac{-1}{4 \tilde{\omega}^5 c_g} \left[ (\tilde{k}^2 - \tilde{\omega}^4)^3 + (2\tilde{k}^2 - \tilde{\omega}^4)^3 + 4(\tilde{k}^2 + \tilde{\omega}^4) - \tilde{k}^4 \tilde{\omega}^4 \right] 
+ \frac{1}{2 \tilde{\omega} c_g} \frac{\left[2 \tilde{k} \tilde{\omega} + \tilde{k}^2 c_g \sech^2 \tilde{k}H \right]^2}{H - c_g^2}.
\end{align}
}
\end{document}